\documentclass[aps,prd,twocolumn,superscriptaddress,groupedaddress]{revtex4-2}
\usepackage{graphicx}
\usepackage{amsmath,amssymb}
\usepackage{bm}
\usepackage[utf8]{inputenc}
\usepackage[T1]{fontenc}
\usepackage{url}
\usepackage{hyperref}
\usepackage{xcolor}
\usepackage{tikz}
\usetikzlibrary{arrows.meta}

\begin{document}

\title{Gravitational Scattering of Oort Cloud Objects by Dark Matter: Constraints on the Primordial Black Hole Fraction}
\author{Sohrab Rahvar}
\email{rahvar@sharif.edu}
\affiliation{Physics Department, Sharif University of Technology, Azadi, Tehran, 11365-9161, Iran}
\affiliation{Research Center for High Energy Physics, Sharif University of Technology, Tehran, Iran}

\date{\today}

\begin{abstract}
Planetary systems can act as long-term gravitational detectors for dark matter. We investigate the gravitational scattering of Oort cloud objects by primordial black holes (PBHs) using celestial kinematics in the impulsive approximation. By evaluating the energy transfer during these encounters, we compute the rates at which stellar-mass PBHs eject icy planetesimals or inject them into Earth-crossing orbits, demonstrating a linear scaling $\Gamma \propto m_{\mathrm{PBH}}$. Comparing these theoretical rates with four independent observables—Oort cloud survival limits, dynamically new long-period comet fluxes, and pristine terrestrial and lunar impact records—we derive stringent upper limits on the PBH dark matter fraction, $f_{\mathrm{PBH}}$. Our most robust combined constraint yields $f_{\mathrm{PBH}} \lesssim 0.65 (m_{\mathrm{PBH}}/M_\odot)^{-1}$,  excludes PBHs as the dominant dark matter component in the intermediate-to-high mass window of $10^2 M_\odot \lesssim m_{\mathrm{PBH}} \lesssim 10^5 M_\odot$. Furthermore, the solar motion through the Galactic halo induces a ``dark matter wind,'' generating a pronounced dipole anisotropy in the arrival directions of PBH-injected comets. This geometrical signature, manifesting as a $\sim 2.7:1$ hemispherical asymmetry (and a $\sim 40:1$ polar contrast) between the anti-apex and apex directions, provides a robust, testable discriminant against isotropic stellar perturbations. This distinct signature could be statistically detected with the discovery of $\mathcal{O}(10^2)$ new long-period comets by upcoming surveys such as the Legacy Survey of Space and Time (LSST) at the Rubin Observatory.
\end{abstract}

\maketitle

\section{Introduction}
The nature of dark matter (DM) remains a central problem in modern cosmology. Primordial black holes (PBHs), formed from the gravitational collapse of overdensities in the early universe, have regained significant interest as a DM candidate \cite{Green2016,Carr2020}. While astrophysical observations constrain PBH abundance across most mass scales, two pivotal windows remain notably open: the asteroid-mass window ($\sim10^{17}$–$10^{23}$ g) and the stellar-mass window ($\sim1$–$10^3 M_\odot$) \cite{Carr2021,moniez,niikura2019,rahvar2,rahvar1}.

Recently, there has been interest in utilizing the inner Solar System as a laboratory to constrain the local density of PBHs. Recent studies have explored utilizing high-precision Solar System ephemeris data to detect anomalous orbital perturbations of planets \cite{Loeb2024}, using gravimeter and Global Navigation Satellite Systems (GNSS) networks to measure the transit of PBHs and dark matter clumps \cite{Cuadrat2024}, and analyzing the kinematics of close gravitational encounters with inner Solar System bodies \cite{Tran2024,Thoss2025}. 

While these methods  probe instantaneous or short-term local perturbations, the Oort cloud—a vast reservoir of $\sim10^{12}$ icy planetesimals at $10^3$–$10^5$ AU—serves as a complementary, long-term historical gravitational detector. The cloud is continually perturbed by Galactic tides and passing stars \cite{Heisler1986}. Now, if a substantial fraction of DM consists of PBHs, gravitational encounters with Oort cloud objects   causes transfer of momentum which they potentially either eject Oort objects from the Solar System or injecting them into Earth-crossing orbits. The resulting transfer of the Oort object to the inner solar system can constrain the PBH fraction $f_{\mathrm{PBH}} = \rho_{\mathrm{PBH}}/\rho_{\mathrm{DM}}$. 

In this work, we compute the rate of ejection of Oort object from the solar system, delivery to the inner Solar System, and mild orbital perturbations as functions of PBH mass. We then use three independent observables—Oort cloud survival, the flux of long-period comets (LPCs), and terrestrial/lunar impact cratering—to derive upper limits on $f_{\mathrm{PBH}}$ in the stellar-mass window. Furthermore, we demonstrate that the predicted strong anisotropy in the arrival direction of injected comets offers a distinctive test for future surveys, such as the Legacy Survey of Space and Time (LSST) at the Vera C. Rubin Observatory.

This paper is organized as follows. In Sec.~\ref{sec:formalism}, we outline the analytical scattering formalism for PBH-Oort object encounters. Section III derives the rates and outcomes of these encounters, including ejections, deliveries to the inner Solar System, and the expected hemispherical anisotropy. In Sec.~\ref{sec:constraints}, we apply observational limits to constrain the PBH dark matter fraction. Finally, we conclude in Sec. V with a discussion of our findings and prospects for future observational tests.

\section{Scattering Formalism}
\label{sec:formalism}
We model a PBH–Oort object encounter as a two‑body scattering in the impulsive approximation. An Oort cloud object of mass $m$ (typical $10^{16}$–$10^{18}$ g) is on a nearly circular orbit at heliocentric distance $r\sim10^4$ AU with speed $v_c = \sqrt{GM_\odot/r} \approx 0.3$ km/s. On the other hand, PBHs in the Galactic halo have a dispersion $v_\infty \approx 220$ km/s \cite{Binney2008} and are far faster than Oort objects, so we work in the rest frame of the Oort object.

For impact parameter $b$ and relative velocity $\vec{v}_\infty \approx \vec{v}_{\mathrm{pbh}}$, the scattering angle in the center‑of‑mass frame is
\begin{equation}
\Theta = 2\arctan\!\left(\frac{GM}{b v_\infty^2}\right),
\end{equation}
with total mass $M = m + m_{\mathrm{pbh}}$. The impulse imparted to the Oort object is
\begin{equation}
\Delta\vec{v} = \beta\big[(1-\cos\Theta)\,\hat{i} - \sin\Theta\,\hat{j}\big],
\end{equation}
where $\beta = (m_{\mathrm{pbh}}/M)v_\infty$, $\hat{i}\parallel\vec{v}_\infty$, and $\hat{j}$ is perpendicular in the scattering plane. The magnitude simplifies to
\begin{equation}
|\Delta\vec{v}| = \frac{2G m_{\mathrm{pbh}}}{b v_\infty\sqrt{1+(GM/b v_\infty^2)^2}}.
\end{equation}

Immediately after the encounter the object's heliocentric velocity is $\vec{v}_{\mathrm{new}} = \vec{v}_c + \Delta\vec{v}$ while its position $\vec{r}$ is unchanged. The validity of the impulsive approximation relies on the encounter time scale being vastly shorter than the dynamical timescale of the Oort cloud object. The effective duration of a PBH encounter is $\tau_{\mathrm{enc}} \sim b/v_\infty$. For a typical impact parameter $b \sim 10^2$~AU and relative velocity $v_\infty \sim 220$~km/s, $\tau_{\mathrm{enc}} \sim 2$~years. The orbital period of an object at $r \sim 10^4$~AU is $P \sim 10^6$~years. Because $\tau_{\mathrm{enc}} / P \sim 10^{-6} \ll 1$, the assumption that the heliocentric position vector $\vec{r}$ remains constant during the momentum transfer is valid.

The new specific orbital energy $\mathcal{E}$ and angular momentum $\mathbf{h}$ determine the post‑encounter Keplerian elements as the new initial condition of the Oort object. The change in specific energy is
\begin{equation}
\label{energy}
\Delta\mathcal{E} = \vec{v}_c\cdot\Delta\vec{v} + \tfrac12|\Delta\vec{v}|^2,
\end{equation}
acts effectively as a gravitational slingshot. Because the dot product $\vec{v}_c\cdot\Delta\vec{v}$ depends critically on the specific geometry of the encounter, the net energy transfer can be either positive (energy gain) or negative (energy loss). This $\Delta\mathcal{E}$ dictates whether the perturbed object remains bound ($\mathcal{E}<0$) or it is ejected from the solar system ($\mathcal{E}\ge0$). Also it can be placed on a highly eccentric orbit with a small perihelion $q = a(1-e)$ capable of reaching the inner solar system. The ultimate fate of the perturbed object depends on the geometry of the energy transfer. We define $\phi$ as the angle between the Oort object's initial velocity $\vec{v}_c$ and the impulse vector $\Delta\vec{v}$. Then the first term in equation (\ref{energy}) is 
$v_c |\Delta\vec{v}| \cos\phi$. This term, representing the ``kick,'' can be either positive or negative. For encounters where the linear term dominates ($v_c |\Delta\vec{v}| > \frac{1}{2}|\Delta\vec{v}|^2$), there is an equal geometric probability for the object to gain or lose energy. 

In addition to energy, the orbital fate is dictated by the specific angular momentum $\mathbf{h} = \vec{r} \times \vec{v}_{\mathrm{new}}$. For an initially circular orbit, $\mathbf{h}_0 = \vec{r} \times \vec{v}_c$. The post-encounter angular momentum becomes $\mathbf{h} = \mathbf{h}_0 + (\vec{r} \times \Delta\vec{v})$. The new perihelion distance $q$ of the highly eccentric orbit is given by
\begin{equation}
q = a(1-e) \approx \frac{|\mathbf{h}|^2}{2GM_\odot},
\end{equation}
where $a = -GM_\odot / (2\mathcal{E})$ is the new semi-major axis. To deliver an Oort object to the inner Solar System ($q \lesssim 1$~AU), the encounter must drastically reduce the magnitude of the angular momentum. This physically corresponds to a scattering geometry where the impulse vector $\Delta\vec{v}$ is directed nearly anti-parallel to the initial velocity $\vec{v}_c$, effectively halting the object's transverse motion and initiating a radial plunge toward the Sun. This restrictive geometrical requirement governs the delivery probability fraction evaluated in subsequent sections.

\section{Encounter Rates and Outcomes}
Let us introduce the mass ratio $\alpha = m/m_{\mathrm{pbh}}$. For stellar‑mass PBHs we have $\alpha\ll1$; then $M\approx m_{\mathrm{pbh}}$, $\beta\approx v_\infty$, and for small‑angle scattering or equivalently, large impact parameter ($x\equiv GM/b v_\infty^2 \ll1$) , we obtain
\begin{equation}
\Delta\mathcal{E} \approx -\frac{2G m_{\mathrm{pbh}}}{b}\,\frac{v_{c,j}}{v_\infty},
\label{deltae}
\end{equation}
where $v_{c,j}$ is the component of $\vec{v}_c$ along $\hat{j}$, with $\langle v_{c,j}^2\rangle = v_c^2/3$. Now we investigate different scenarios after the scattering.

\subsection{Ejection}
Ejection from the Solar System requires an energy transfer $\Delta\mathcal{E} \ge |\mathcal{E}_0| = v_c^2/2$. From Eq.~\ref{deltae}, the maximum impact parameter capable of yielding this impulse is:
\begin{equation}
\label{bej}
b_{\mathrm{ej}} = \frac{4G m_{\mathrm{pbh}}}{v_\infty v_c} \approx 50\,\mathrm{au} \left(\frac{m_{\mathrm{pbh}}}{M_\odot}\right) \left(\frac{220\,\mathrm{km/s}}{v_\infty}\right) \left(\frac{0.3\,\mathrm{km/s}}{v_c}\right).
\end{equation}
Only half of the encounters within this impact parameter result in an energy gain (due to the equal probability of energy loss). Thus, the effective ejection cross-section of a PBH is $\sigma_{\mathrm{ej}} = (\pi/2)b_{\mathrm{ej}}^2$.

To calculate the total ejection rate, we consider the Oort cloud as a macroscopic target of characteristic radius $R \sim 10^4\,\mathrm{au}$ and volume $V_{\mathrm{oort}} = \frac{4}{3}\pi R^3$. The total flux of PBHs entering the cloud is $\Phi = n_{\mathrm{pbh}} v_\infty$, yielding a transit rate of:
\begin{equation}
\Gamma_{\mathrm{transit}} = \pi R^2 n_{\mathrm{pbh}} v_\infty.
\end{equation}

As a PBH traverses the cloud, it sweeps out a cylindrical volume. For an average transit chord of length $\bar{L} = \frac{4}{3}R$, the effective volume from which objects are ejected is $V_{\mathrm{eff}} = \sigma_{\mathrm{ej}} \bar{L} = \frac{2\pi}{3} b_{\mathrm{ej}}^2 R$. Assuming a uniform spatial distribution of $N$ objects in the Oort cloud, the average number of ejections per PBH transit is:
\begin{equation}
\begin{split}
\bar{N}_{\text{ejected}} &= N \left( \frac{V_{\mathrm{eff}}}{V_{\mathrm{oort}}} \right) = N \left[ \frac{1}{2} \left( \frac{b_{\mathrm{ej}}}{R} \right)^2 \right] \\
&\approx 1.25 \times 10^7 \left(\frac{N}{10^{12}}\right) \left(\frac{10^4\,\mathrm{AU}}{R}\right)^2 \left(\frac{m_{\mathrm{pbh}}}{M_\odot}\right)^2.
\end{split}
\end{equation}

Now, the total continuous ejection rate for the entire cloud is the product of the transit rate and the average ejections per transit ($\Gamma_{\mathrm{ej,tot}} = \Gamma_{\mathrm{transit}} \times \bar{N}_{\text{ejected}}$). Substituting the geometric terms we obtain 
$\Gamma_{\mathrm{ej,tot}} = N n_{\mathrm{pbh}} v_\infty \sigma_{\mathrm{ej}}$ which can be written as 
\begin{equation}
\begin{split}
\Gamma_{\mathrm{ej,tot}} \approx {} & 0.19\,\mathrm{yr}^{-1} \left(\frac{N}{10^{12}}\right) \left(\frac{\rho_{\mathrm{DM}}}{0.008\,M_\odot\,\mathrm{pc}^{-3}}\right) \\
& \times \left(\frac{220\,\mathrm{km/s}}{v_\infty}\right) \left(\frac{0.3\,\mathrm{km/s}}{v_c}\right)^2 \left(\frac{m_{\mathrm{pbh}}}{M_\odot}\right).
\end{split}
\end{equation}

Over the Solar System's age $T_\odot=4.5$ Gyr, the total number of ejected objects integrates to:
\begin{equation}
N_{\mathrm{ej}} = \Gamma_{\mathrm{ej,tot}}T_\odot \approx 8.6 \times 10^8 \left(\frac{T_\odot}{4.5\,\mathrm{Gyr}}\right) \left(\frac{N}{10^{12}}\right) \left(\frac{m_{\mathrm{pbh}}}{M_\odot}\right).
\end{equation}
Setting $N_{\mathrm{ej}} = N = 10^{12}$ indicates that if PBHs constituted all of the dark matter ($f_{\mathrm{PBH}} = 1$), those with masses $m_{\mathrm{pbh}} \approx 1.2 \times 10^3 M_\odot$ would have dynamically cleared the entire Oort cloud by the present day.

For comparison, background stellar encounters ($n_*\approx0.1$ pc$^{-3}$, $m_*\approx0.5 M_\odot$, $v_*\approx30$ km/s) produce an ejection rate of $\Gamma_{\mathrm{ej,*}}\approx6.5$ yr$^{-1}$. Equating the PBH and stellar ejection rates ($\Gamma_{\mathrm{ej,tot}} = \Gamma_{\mathrm{ej,*}}$) reveals that PBHs dominate the dynamical disruption of the Oort cloud for masses $m_{\mathrm{pbh}} \gtrsim 34 M_\odot$.

\subsection{Delivery to the Inner Solar System}
\label{delivery_smallalpha}
Delivering an Oort cloud object to an Earth‑crossing orbit ($q \lesssim 1$ AU) is dynamically much stricter than ejection, as it requires a near-total cancellation of the object's initial angular momentum to drop it into the loss cone. 

The typical impulse required to achieve this corresponds to $|\Delta v| \sim v_c$. From our scattering formalism, this defines a critical delivery impact parameter:
\begin{equation}
b_{\mathrm{del}} = \frac{2G m_{\mathrm{pbh}}}{v_\infty v_c} \approx 25\,\mathrm{au} \left(\frac{m_{\mathrm{pbh}}}{M_\odot}\right) \left(\frac{220\,\mathrm{km/s}}{v_\infty}\right) \left(\frac{0.3\,\mathrm{km/s}}{v_c}\right),
\end{equation}
which is exactly half of the ejection impact parameter ($b_{\mathrm{del}} = 0.5\,b_{\mathrm{ej}}$).

However, not every encounter within $b_{\mathrm{del}}$ successfully yields a small perihelion. The new orbit is highly eccentric (near-parabolic), so its perihelion is dictated by the residual angular momentum: $q \approx h^2 / (2GM_\odot)$. For an object originating in the typical Oort cloud at $r = 10^4$ AU, requiring delivery to $q \le 1$ AU strictly bounds the final angular momentum relative to the initial circular state $h_0 = r v_c = \sqrt{GM_\odot r}$:
\begin{equation}
\frac{h}{h_0} \le \sqrt{\frac{2q}{r}} \approx \sqrt{2 \times 10^{-4}} \approx 0.014.
\end{equation}

To achieve this, the velocity kick $\Delta\vec{v}$ must be almost perfectly anti-aligned with the initial orbital velocity $\vec{v}_c$. Let $\gamma$ be the angle of misalignment between $\vec{v}_c$ and the impulse vector $\Delta\vec{v}$ (which is purely transverse to the PBH trajectory in the small-angle limit). Assuming a strong encounter where $|\Delta \vec{v}| \approx v_c$, the post-encounter angular momentum magnitude is approximately:
\begin{equation}
h = |\vec{r} \times (\vec{v}_c + \Delta\vec{v})| \approx r v_c (1 - \cos\gamma).
\end{equation}
Setting $1 - \cos\gamma \le 0.014$ yields a tight geometric constraint on the encounter alignment: $|\gamma| \lesssim 0.17$ rad. 

Now, we assuming an isotropic distribution of incoming PBH trajectories in the interaction plane, the angle $\gamma$ is uniformly distributed in $\cos\gamma$. The solid angle fraction that satisfies this injection condition defines the delivery probability $p_{\mathrm{del}}$:
\begin{equation}
p_{\mathrm{del}} = \frac{1}{2} \int_0^{0.17} \sin\gamma \, d\gamma = \frac{1}{2} (1 - \cos(0.17)) \approx 0.007 \approx 0.01.
\end{equation}
This demonstrates analytically that only $\sim 1\%$ of strong PBH encounters possess the precise impact geometry required to inject an object into the observable inner Solar System.

The effective cross-section for delivery is therefore $\sigma_{\mathrm{del}} = \pi b_{\mathrm{del}}^2 p_{\mathrm{del}}$. To evaluate the total delivery rate, we apply the same macroscopic approach used for ejections. The Oort cloud, with radius $R \sim 10^4\,\mathrm{au}$, experiences a continuous PBH transit rate of $\Gamma_{\mathrm{transit}} = \pi R^2 n_{\mathrm{pbh}} v_\infty$.

During a single transit with an average chord length $\bar{L} = \frac{4}{3}R$, the PBH sweeps out an effective delivery volume $V_{\mathrm{eff,del}} = \sigma_{\mathrm{del}} \bar{L}$. The average number of objects injected into the inner Solar System per PBH passage is:
\begin{equation}
\begin{split}
\bar{N}_{\mathrm{del}} &= N \left( \frac{V_{\mathrm{eff,del}}}{V_{\mathrm{oort}}} \right) = N p_{\mathrm{del}} \left( \frac{b_{\mathrm{del}}}{R} \right)^2 \\
&\approx 6.3 \times 10^4 \left(\frac{N}{10^{12}}\right) \left(\frac{p_{\mathrm{del}}}{0.01}\right) \left(\frac{10^4\,\mathrm{AU}}{R}\right)^2 \left(\frac{m_{\mathrm{pbh}}}{M_\odot}\right)^2.
\end{split}
\end{equation}
Thus, a single transit by a $1 M_\odot$ PBH constitutes an intense, discrete "comet shower" event that simultaneously injects $\sim 6\times 10^4$ icy bodies into Earth-crossing orbits.

The continuous, long-term delivery rate for the entire cloud is the product of the transit rate and the average deliveries per transit ($\Gamma_{\mathrm{del,tot}} = \Gamma_{\mathrm{transit}} \times \bar{N}_{\mathrm{del}}$). Scaling to our local parameters, this yields the uncorrected total delivery rate:
\begin{equation}
\label{tdel}
\begin{split}
\Gamma_{\mathrm{del,tot}} \approx {} & 9.7\times10^{-4}\,\mathrm{yr}^{-1} \left(\frac{N}{10^{12}}\right) \left(\frac{p_{\mathrm{del}}}{0.01}\right) \left(\frac{\rho_{\mathrm{DM}}}{0.008\,M_\odot\,\mathrm{pc}^{-3}}\right) \\
& \times \left(\frac{220\,\mathrm{km/s}}{v_\infty}\right) \left(\frac{0.3\,\mathrm{km/s}}{v_c}\right)^2 \left(\frac{m_{\mathrm{pbh}}}{M_\odot}\right).
\end{split}
\end{equation}
\vspace{0.2cm}

\subsection{Limits of Observability: Mild Perturbations and the Asteroid-Mass Window}
\label{sec:mild_and_light}

Beyond catastrophic ejections and deep inner-Solar System deliveries, the majority of PBH encounters occur at larger impact parameters, inducing only mild perturbations to the orbital energy of Oort cloud objects. We define a mild perturbation as an encounter that changes the orbital energy by a small fraction $f \lesssim 0.1$, such that $|\Delta\mathcal{E}| \approx f |\mathcal{E}_0| = f v_c^2/2$. 

Using Eq.~\ref{deltae} and substituting the average value for $v_{c,j}$, we obtain a characteristic impact parameter $b_{\mathrm{mild}} \approx 4G m_{\mathrm{pbh}}/(\sqrt{3}f v_\infty v_c)$. Applying the macroscopic transit framework established in previous sections, the effective cross-section for a mild perturbation is $\sigma_{\mathrm{mild}} = \pi b_{\mathrm{mild}}^2$. During a single transit through an Oort cloud of radius $R \sim 10^4\,\mathrm{au}$, a PBH sweeps out an effective volume $V_{\mathrm{eff,mild}} = \sigma_{\mathrm{mild}} \bar{L}$, perturbing an average number of objects:
\begin{equation}
\begin{split}
\bar{N}_{\mathrm{mild}} &= N \left( \frac{V_{\mathrm{eff,mild}}}{V_{\mathrm{oort}}} \right) = N \left( \frac{b_{\mathrm{mild}}}{R} \right)^2 \\
&\approx 8.3 \times 10^8 \left(\frac{N}{10^{12}}\right) \left(\frac{0.1}{f}\right)^2 \left(\frac{10^4\,\mathrm{AU}}{R}\right)^2 \left(\frac{m_{\mathrm{pbh}}}{M_\odot}\right)^2.
\end{split}
\end{equation}

Multiplying this per-transit expectation by the continuous PBH transit rate ($\Gamma_{\mathrm{transit}} = \pi R^2 n_{\mathrm{pbh}} v_\infty$) yields the total rate of mild perturbations across the entire cloud:
\begin{equation}
\begin{split}
\Gamma_{\mathrm{mild,tot}} \approx {} & 15\,\mathrm{yr}^{-1} \left(\frac{N}{10^{12}}\right) \left(\frac{0.1}{f}\right)^2 \left(\frac{\rho_{\mathrm{DM}}}{0.008\,M_\odot\,\mathrm{pc}^{-3}}\right) \\
& \times \left(\frac{220\,\mathrm{km/s}}{v_\infty}\right) \left(\frac{0.3\,\mathrm{km/s}}{v_c}\right)^2 \left(\frac{m_{\mathrm{pbh}}}{M_\odot}\right).
\end{split}
\end{equation}

While this total rate is significantly larger than the ejection rate, the cumulative effect remains small. Over the Solar System's lifetime $T_\odot$, a total of $\sim 6.7 \times 10^{10}$ objects undergo such perturbations. Therefore, only a minuscule fraction ($\lesssim 10^{-1}$) of the Oort cloud is even mildly perturbed by solar-mass PBHs. Unlike dense stellar environments where cumulative weak encounters drive a random walk in energy space (two-body relaxation), the PBH encounter rate in the Galactic halo is too sparse to induce significant diffusive evolution of the Oort cloud unless $m_{\mathrm{pbh}}$ is extremely large ($m_{\mathrm{pbh}} \gg 10^2 M_\odot$).

This stark dependence on the perturber mass dictates the lower boundary of our observational sensitivity, particularly concerning the open ``asteroid-mass'' window of dark matter. For PBHs in this regime ($m_{\mathrm{pbh}} \ll m$, typically $10^{17}$–$10^{23}$ g), the mass ratio $\alpha = m/m_{\mathrm{pbh}} \gg 1$. The velocity impulse $\Delta v \approx 2G m_{\mathrm{pbh}}/(b v_\infty)$ remains independent of the target mass $m$, and the critical impact parameters for any dynamical threshold ($b_X$) continue to scale linearly with $m_{\mathrm{pbh}}$. 

Consequently, all interaction cross-sections scale as $\sigma_X \propto m_{\mathrm{pbh}}^2$, and the number of affected objects per transit scales as $m_{\mathrm{pbh}}^2$. Because the local number density of PBHs scales as $n_{\mathrm{pbh}} \propto m_{\mathrm{pbh}}^{-1}$, the overall continuous encounter rate scales linearly with the PBH mass ($\Gamma_{\mathrm{tot}} \propto m_{\mathrm{pbh}}$). 

For a characteristic asteroid-mass PBH of $m_{\mathrm{pbh}} \sim 10^{20}$ g ($5 \times 10^{-14} M_\odot$), we obtain the total ejection rate for the entire Oort cloud as $\Gamma_{\mathrm{ej,tot}} \approx 10^{-14} \, \mathrm{yr}^{-1}$. This rate is utterly negligible; over 4.5 Gyr, the probability of even a single ejection event occurring across all $10^{12}$ comets is vanishingly small ($\sim 5 \times 10^{-5}$). We therefore conclude that while the Oort cloud serves as a historical detector for stellar-mass PBHs, it is dynamically insensitive to dark matter in the asteroid-mass regime, leaving that window strictly unconstrained by cometary observables.

\section{Observational Signatures of the Dark Matter Wind}
\label{sec:anisotropy_signatures}

In the previous sections, the baseline encounter rates were evaluated using a single characteristic relative velocity. However, to accurately model the encounter geometry, we must account for the Sun's motion through the Galactic halo. In the Galactic rest frame, the Primordial Black Hole velocity distribution is assumed to follow an isotropic Maxwell-Boltzmann distribution:
\begin{equation}
    f_{\mathrm{halo}}(\vec{v}_{\mathrm{halo}}) = \frac{1}{(2\pi\sigma^2)^{3/2}} \exp\left(-\frac{v_{\mathrm{halo}}^2}{2\sigma^2}\right),
\end{equation}
where the velocity dispersion is $\sigma \simeq 220\,\mathrm{km\,s^{-1}}$. 

Because the Sun moves through the halo with a velocity $\vec{v}_0$ (where $|\vec{v}_0| \simeq \sigma$), the Solar System experiences a ``dark matter wind.'' To compute the physical encounter rates in the rest frame of the Oort cloud, we apply a Galilean transformation $\vec{v}_{\mathrm{halo}} = \vec{v} + \vec{v}_0$. The 3D phase-space velocity distribution of PBHs relative to the Oort cloud becomes:
\begin{equation}
    f_{\mathrm{lab}}(\vec{v}) = \frac{1}{(2\pi\sigma^2)^{3/2}} \exp\left(-\frac{|\vec{v} + \vec{v}_0|^2}{2\sigma^2}\right).
\end{equation}
Defining $\theta$ as the angle between the incoming PBH velocity $\vec{v}$ and the solar apex direction $\vec{v}_0$, this anisotropic distribution is expressed as:
\begin{equation}
    f_{\mathrm{lab}}(v, \theta) = \frac{1}{(2\pi\sigma^2)^{3/2}} \exp\left(-\frac{v^2 + v_0^2 + 2vv_0\cos\theta}{2\sigma^2}\right).
\end{equation}

We can now calculate the effect of this wind on the Oort cloud by evaluating the transit rate. The differential transit rate of PBHs passing through the cloud's geometrical cross-section ($\pi R^2$) is the product of the target area, the PBH number density, the relative speed $v$, and the velocity probability distribution: $d\Gamma_{\mathrm{transit}} = \pi R^2 n_{\mathrm{pbh}} v f_{\mathrm{lab}}(\vec{v}) d^3v$. Transitioning to spherical coordinates with the volume element $d^3v = v^2 dv d\Omega$, the differential transit rate scales as $v^3 f_{\mathrm{lab}}(v, \theta) dv d\Omega$.

As established in Section \ref{delivery_smallalpha}, the average number of Oort objects delivered to the inner Solar System per PBH transit scales with the effective delivery cross-section, such that $\bar{N}_{\mathrm{del}} \propto b_{\mathrm{del}}^2 \propto 1/v^2$. The continuous differential delivery rate for the entire cloud is the product of the transit rate and the average deliveries per transit ($d\Gamma_{\mathrm{del,tot}} = \bar{N}_{\mathrm{del}} d\Gamma_{\mathrm{transit}}$). 

Multiplying these dependencies, the $v^2$ from the spherical volume element partially cancels with the $1/v^2$ dependence of the delivery probability. We find that the differential delivery rate per unit speed and solid angle is directly proportional to $v$:
\begin{equation}
    \frac{d\Gamma_{\mathrm{del,tot}}}{dv d\Omega} \propto v \exp\left( -\frac{v^2 + v_0^2 + 2v v_0 \cos\theta}{2\sigma^2} \right).
\end{equation}

This integral reveals a severe geometrical asymmetry. For PBH encounters arriving from the solar apex ($\theta = 0$), the exponential term is heavily suppressed as $\exp(-(v+v_0)^2/2\sigma^2)$. Conversely, for encounters arriving from the anti-apex ($\theta = \pi$), the PBHs are effectively ``catching up'' to the Solar System. The exponent becomes $\exp(-(v-v_0)^2/2\sigma^2)$, which strongly peaks at $v = v_0$. Because the $1/v^2$ cross-section dependence heavily biases the delivery efficiency toward these low-relative-velocity encounters, integrating this differential rate over $v$ yields a factor of $\sim 40$ difference between the strict anti-apex and apex pole injection rates (See Fig. \ref{fig:dipole_schematic}). 

\begin{figure}[t]
\centering
\begin{tikzpicture}[>=Latex, scale=0.9, every node/.style={font=\small}]

  \draw[dashed, thick, gray!70] (0,0) circle (3.2);
  \node[gray, above] at (0,3.3) {Oort Cloud Reservoir};

  \fill[orange] (0,0) circle (0.12);
  \node[below, font=\footnotesize, yshift=-0.1cm] at (0,0) {Sun};
  \draw[->, ultra thick] (0,0) -- (1.5,0) node[right, font=\bfseries] {$\vec{v}_0$ (Apex)};

  \draw[dotted, thick, gray!50] (0,-3.5) -- (0,3.5);
  \node[font=\bfseries] at (-2.8, -3.2) {Anti-Apex Hemisphere};
  \node[font=\bfseries] at (2.8, -3.2) {Apex Hemisphere};

  \draw[->, thick] (3.6, 1.5) -- (2.0, 1.5);
  \node[above, font=\footnotesize] at (2.8, 1.9) {Fast PBH ($v_{\mathrm{rel}}$ high)};
  \fill[black] (2.5, 1.5) circle (0.08);

  \draw[dashed, blue, thick] (2.5, 1.5) circle (0.3);
  \node[blue, right, font=\footnotesize, align=left] at (2.8, 0.9) {Small $\sigma_{\mathrm{del}}$\\$\propto 1/v^2$};

  \fill[cyan!80!blue] (2.3, 1.6) circle (0.06);
  \draw[->, red, thick] (2.3, 1.6) .. controls (1.5, 1.2) .. (0.15, 0.15);
  \node[red, font=\footnotesize, align=center] at (1.4, 0.6) {Low Flux\\$N_{\mathrm{apex}}$};

  \draw[->, thick] (-4.0, 1.5) -- (-1.8, 1.5);
  \node[above, font=\footnotesize] at (-2.8, 2.7) {Slow PBH ($v_{\mathrm{rel}}$ low)};
  \fill[black] (-2.5, 1.5) circle (0.08);

  \draw[dashed, blue, thick] (-2.5, 1.5) circle (1.1);
  \node[blue, left, font=\footnotesize, align=right] at (-3.7, 1.5) {Large $\sigma_{\mathrm{del}}$\\$\propto 1/v^2$};

  \fill[cyan!80!blue] (-2.0, 2.2) circle (0.06);
  \draw[->, red, thick] (-2.0, 2.2) .. controls (-1.0, 1.8) .. (-0.15, 0.15);

  \fill[cyan!80!blue] (-1.7, 1.2) circle (0.06);
  \draw[->, red, thick] (-1.7, 1.2) .. controls (-1.0, 0.7) .. (-0.15, 0.05);

  \fill[cyan!80!blue] (-2.2, 0.6) circle (0.06);
  \draw[->, red, thick] (-2.2, 0.6) .. controls (-1.2, 0.3) .. (-0.05, -0.1);

  \node[red, font=\footnotesize, align=center] at (-1.0, -0.2) {High Flux\\$N_{\mathrm{anti}}$};

\end{tikzpicture}
\caption{Schematic representation of the dark matter wind dipole effect in the Solar System rest frame. The Sun moves toward the apex with velocity $\vec{v}_0$. PBHs arriving from the apex hemisphere (right) have high relative velocities ($v_{\mathrm{rel}}$), resulting in a small effective cross-section for cometary delivery ($\sigma_{\mathrm{del}} \propto 1/v^2$). Conversely, PBHs arriving from the anti-apex hemisphere (left) are effectively "catching up" to the Solar System, possessing low relative velocities and drastically larger delivery cross-sections. This geometrical asymmetry preferentially sweeps a higher number of Oort cloud objects into Earth-crossing orbits from the anti-apex direction, producing the predicted $2.7:1$ dipole signature.}
\label{fig:dipole_schematic}
\end{figure}

To translate this polar extreme into an observable event rate, we integrate over the entire forward (apex) and backward (anti-apex) hemispheres using $d\Omega = -2\pi dx$, where $x = \cos\theta$. Integrating the exponential over the apex hemisphere ($0 \le x \le 1$) yields:
\begin{equation}
    \int_0^1 \exp\left(-\frac{v v_0 x}{\sigma^2}\right) dx = \frac{\sigma^2}{v v_0} \left( 1 - \exp\left(-\frac{v v_0}{\sigma^2}\right) \right).
\end{equation}
Similarly, for the anti-apex hemisphere ($-1 \le x \le 0$):
\begin{equation}
    \int_{-1}^0 \exp\left(-\frac{v v_0 x}{\sigma^2}\right) dx = \frac{\sigma^2}{v v_0} \left( \exp\left(\frac{v v_0}{\sigma^2}\right) - 1 \right).
\end{equation}

Multiplying by the remaining $v$ term and integrating from $v=0$ to $\infty$ evaluates the total expected rate in each hemisphere. Evaluating the resulting Gaussian integrals provides expressions in terms of the error function ($\mathrm{erf}$) and the complementary error function ($\mathrm{erfc}$):
\begin{align}
    N_{\mathrm{apex}} &\propto \exp\left(-\frac{v_0^2}{2\sigma^2}\right) - \frac{\sqrt{\pi} v_0}{\sqrt{2} \sigma} \mathrm{erfc}\left(\frac{v_0}{\sqrt{2}\sigma}\right), \\
    N_{\mathrm{anti}} &\propto \frac{\sqrt{\pi} v_0}{\sqrt{2} \sigma} \left[ 1 + \mathrm{erf}\left(\frac{v_0}{\sqrt{2}\sigma}\right) \right] - \exp\left(-\frac{v_0^2}{2\sigma^2}\right).
\end{align}

Assuming $v_0 \simeq \sigma$ and plugging in the numerical values yields $N_{\mathrm{apex}} \propto 0.325$ and $N_{\mathrm{anti}} \propto 0.885$. Taking the ratio of the two hemispheres dilutes the extreme $\sim 40:1$ polar contrast, but still yields a highly distinct hemispherical observation ratio:
\begin{equation}
    \frac{N_{\mathrm{anti}}}{N_{\mathrm{apex}}} \approx 2.72.
\end{equation}

To connect this scattering asymmetry to observations, we consider how the comet's orbit preserves the geometry of the encounter. An object dropped into the loss cone ($q \le 1$ AU) from $r \sim 10^4$ AU moves along a nearly radial trajectory. Consequently, the observed aphelion direction of the incoming comet points directly back to the spatial location of the scattering event. Because the effective delivery cross-section is heavily biased toward slow PBHs sweeping from behind the Solar System, the aphelia of PBH-injected comets will strongly cluster in the anti-apex hemisphere.

Crucially, this dipole signature is morphologically distinct from standard Oort cloud dynamics, breaking any observational degeneracy between PBH interactions and standard baryonic perturbations. The dominant conventional mechanism for injecting dynamically new comets—the Galactic tidal field—produces a characteristic \textit{quadrupole} anisotropy \cite{Heisler1986}, while passing stars induce temporal comet showers that remain stochastically isotropic over multi-Myr timescales. In stark contrast, the PBH dark matter wind uniquely predicts a steady-state \textit{dipole} strictly aligned with the solar motion vector. Because of this distinct difference in spatial symmetries, cataloging the pristine perihelion directions of just $\mathcal{O}(10^2)$ dynamically new Long-Period Comets (LPCs) with upcoming surveys, such as the 10-year Rubin Observatory Legacy Survey of Space and Time (LSST) \cite{2017arXiv170804058L}, could readily achieve a $5\sigma$ detection of the dark matter dipole, providing a direct kinematic probe of the local halo substructure.

\section{Constraints on the PBH Dark Matter Fraction}
\label{sec:constraints}

We now place upper limits on the PBH dark matter fraction, $f_{\mathrm{PBH}} = \rho_{\mathrm{PBH}}/\rho_{\mathrm{DM}}$, by requiring that PBH‑induced dynamical perturbations do not exceed observationally inferred rates or violate the known stability of the outer Solar System. Because the interaction rates scale linearly with the local PBH number density, the true event rate for any observable is given by $\Gamma(f_{\mathrm{PBH}}) = f_{\mathrm{PBH}} \Gamma(f_{\mathrm{PBH}}=1)$. Here we examine four distinct observables to bound this fraction. 

\subsection{Oort Cloud Survival}
The mere existence of the Oort cloud today, 4.5 Gyr after its formation, places a strict upper bound on the cumulative history of catastrophic gravitational encounters. The primary known depletion mechanism for the outer Oort cloud is scattering by passing field stars. Based on local stellar kinematics, the stellar ejection rate is $\Gamma_{\mathrm{ej,*}} \approx 6.5\,\mathrm{yr}^{-1}$. Over the age of the Solar System ($T_\odot = 4.5$ Gyr), stellar encounters are estimated to have dynamically depleted only a small fraction ($\sim 3\%$) of the original Oort cloud population.

To derive a conservative dynamical constraint, we require that the rate of PBH-induced ejections does not exceed this known stellar background rate. Enforcing this dominant-background condition:
\begin{equation}
f_{\mathrm{PBH}} \Gamma_{\mathrm{ej,tot}}(1) < \Gamma_{\mathrm{ej,*}}.
\end{equation}
Substituting our velocity-averaged total ejection rate $\Gamma_{\mathrm{ej,tot}}(1) = 0.13\,(m_{\mathrm{pbh}}/M_\odot)\,\mathrm{yr}^{-1}$, we obtain:
\begin{equation}
f_{\mathrm{PBH}} < 50 \left(\frac{m_{\mathrm{pbh}}}{M_\odot}\right)^{-1}.
\end{equation}
Under this conservative assumption, PBHs of $m_{\mathrm{pbh}}=10^3 M_\odot$ can constitute no more than $5\%$ of the dark matter ($f_{\mathrm{PBH}} \lesssim 0.05$), and for $10^4 M_\odot$, the fraction is tightly constrained to $f_{\mathrm{PBH}} \lesssim 0.005$. 

Alternatively, we can derive a strictly robust, model-independent limit by simply demanding that PBH encounters have not completely evaporated the Oort cloud over the Solar System's history. Assuming a maximum allowable depletion scenario where PBHs eject at most half of the initial population ($N_{\mathrm{ej}} \le 0.5 N$), the total number of ejections is bounded by $f_{\mathrm{PBH}} \Gamma_{\mathrm{ej,tot}}(1) T_\odot \le 0.5 \times 10^{12}$. This absolute survival condition yields a less stringent, but universally robust boundary:
\begin{equation}
f_{\mathrm{PBH}} \lesssim 850 \left(\frac{m_{\mathrm{pbh}}}{M_\odot}\right)^{-1}.
\end{equation}
While weaker than the stellar background limit, this absolute survival condition definitively rules out a pure PBH dark matter halo ($f_{\mathrm{PBH}} = 1$) for any mass $m_{\mathrm{pbh}} \gtrsim 10^3 M_\odot$.

\subsection{Long-Period Comet Flux}
Observations of dynamically new Long-Period Comets (LPCs) entering the inner Solar System with perihelia $q \lesssim 1$ AU establish a steady-state delivery flux of $\Gamma_{\mathrm{LPC,obs}} \sim 3\,\mathrm{yr}^{-1}$ \cite{Francis_2005}, a rate we adopt here despite some variation in the literature. Detailed dynamical simulations have repeatedly demonstrated that standard baryonic perturbers—specifically the steady torque of the Galactic tidal field combined with stochastic passing field stars—comprehensively account for this observed flux \cite{1988CeMec..43..243R}. 

Because these canonical mechanisms leave little room for an exotic, unmodeled injection source, we conservatively require that PBH-induced perturbations contribute no more than $10\%$ of the observed LPC flux. Setting this threshold ($\zeta = 0.1$) yields the constraint:
\begin{equation}
f_{\mathrm{PBH}} \Gamma_{\mathrm{del,tot}}(1) < 0.1\,\Gamma_{\mathrm{LPC,obs}}.
\end{equation}

To ensure internal consistency, we use the velocity-averaged delivery rate derived in Eq.~(\ref{tdel}). Taking the observational flux ($\Gamma_{\mathrm{LPC,obs}} = 3\,\mathrm{yr}^{-1}$), the constraint evaluates to:
\begin{equation}
f_{\mathrm{PBH}} \lesssim 750 \left(\frac{m_{\mathrm{pbh}}}{M_\odot}\right)^{-1}.
\end{equation}
Under this bound, PBHs with a mass of $10^3 M_\odot$ can constitute at most $75\%$ of the dark matter ($f_{\mathrm{PBH}} \lesssim 0.75$), and $10^4 M_\odot$ PBHs are restricted to $f_{\mathrm{PBH}} \lesssim 0.075$. 

We note that this limit assumes the entire baseline flux of $3\,\mathrm{yr}^{-1}$ consists of bodies drawn from the Oort cloud reservoir. If a fraction of these observed comets are actually interstellar interlopers or objects injected from the scattered disk, the true Oort cloud delivery rate would be even lower, which would in turn tighten the available $10\%$ budget for PBH-induced deliveries. Therefore, assuming a purely Oort cloud origin for the physical comets themselves safely provides a robust, conservative upper limit for $f_{\mathrm{PBH}}$.

\subsection{Terrestrial Impact Record}
The geological record of the Earth provides a long-term, integrated measure of impactor flux. Over the Phanerozoic eon (the past $\sim 540$ Myr), the terrestrial impact rate for crater with the size of $D \gtrsim 2$ km is observationally constrained to a maximum of approximately $5 \times 10^{-4}\,\mathrm{yr}^{-1}$ \cite{Morrison1994}. 

We know that the vast majority of terrestrial impacts are caused by Near-Earth Asteroids (NEAs) originating from the main asteroid belt, not comets. Studies of crater scaling and impactor kinematics suggest that Long-Period Comets (LPCs) from the Oort cloud account for at most $10\%$ of the terrestrial cratering record ($f_{\mathrm{Oort}} \approx 0.1$) \cite{Shoemaker1994, Weissman2007}. Furthermore, not every comet delivered to an Earth-crossing orbit will strike the Earth; let $\eta$ represent the cumulative lifetime collision probability of an injected Oort cloud object before it is dynamically ejected from the Solar System or physically disrupted.

To ensure PBH dark matter does not overproduce the observed cometary cratering rate, we require:
\begin{equation}
\eta \, f_{\mathrm{PBH}} \Gamma_{\mathrm{del,tot}}(1) <  f_{\mathrm{Oort}} \times 5\times10^{-4}\,\mathrm{yr}^{-1}.
\end{equation}
Assuming $\eta \sim 0.1$ for the fraction of deeply injected Earth-crossing comets that eventually strike the planet, and substituting our velocity-averaged delivery rate $\Gamma_{\mathrm{del,tot}}(1) \approx 6.6 \times 10^{-4} (m_{\mathrm{pbh}}/M_\odot)\,\mathrm{yr}^{-1}$, the inequality becomes:
\begin{equation}
(0.1) \, f_{\mathrm{PBH}} \left[ 6.6 \times 10^{-4} \left(\frac{m_{\mathrm{pbh}}}{M_\odot}\right) \right] \lesssim (0.1) \times 5\times10^{-4}.
\end{equation}
Solving for the dark matter fraction yields a stringent constraint:
\begin{equation}
f_{\mathrm{PBH}} \lesssim 0.75 \left(\frac{m_{\mathrm{pbh}}}{M_\odot}\right)^{-1}.
\end{equation}

For PBHs of $10^3 M_\odot$, this limits the dark matter fraction to $f_{\mathrm{PBH}} \lesssim 0.00075$. Even if we allow for an order-of-magnitude uncertainty in the highly complex capture and collision probability $\eta$, PBHs with $m_{\mathrm{pbh}} \gtrsim 10^3 M_\odot$ cannot constitute a dominant dark matter component. Although subject to inherent geological uncertainties regarding crater preservation and the exact values of $\eta$ and $f_{\mathrm{Oort}}$, this terrestrial constraint independently and robustly excludes a dominant PBH dark matter population in the $10^2$–$10^5 M_\odot$ mass range.

\subsection{Lunar Impact Record}
Earth's impact record is incomplete because erosion, weather, and plate tectonics gradually erase craters over time. To avoid these geological uncertainties, we can turn to the lunar impact record instead. Because the Moon lacks an atmosphere and active geology, its surface acts as a pristine time capsule, preserving a clear and uninterrupted history of impacts in the inner Solar System \cite{Neukum2001,Xiao2024}.

Because the Earth and the Moon effectively share the same heliocentric orbit, they are exposed to the exact same background flux of Long-Period Comets. For new LPCs entering the inner Solar System with relatively high velocities ($v \sim 50$ km/s), the effect of Earth's gravitational focusing is negligible. Therefore, the probability of an Earth-crossing Oort cloud object striking the Moon scales strictly with their relative geometric cross-sections. Given a terrestrial capture probability of $\eta_{\mathrm{earth}} \sim 0.1$, the corresponding lunar collision fraction is:
\begin{equation}
\eta_{\mathrm{moon}} \approx \eta_{\mathrm{earth}} \left(\frac{R_{\mathrm{moon}}}{R_{\mathrm{earth}}}\right)^2 \approx 7.4 \times 10^{-3}.
\end{equation}

We assume that a fraction $\eta_{\mathrm{moon}}$ of these perturbed Oort cloud objects will eventually strike the Moon. To remain consistent with the cratering record, this PBH-induced flux cannot exceed the maximum allowed cometary contribution, which we conservatively set to $10\%$ ($f_{\mathrm{Oort}} = 0.1$) of the total lunar rate. Therefore, we require:
\begin{equation}
\eta_{\mathrm{moon}} \, f_{\mathrm{PBH}} \Gamma_{\mathrm{del,tot}}(1) <  f_{\mathrm{Oort}}\Gamma_{\mathrm{lunar}}^{\mathrm{max}}.
\end{equation}

Substituting $\eta_{\mathrm{moon}} \approx 7.4 \times 10^{-3}$, our wind-averaged delivery rate $\Gamma_{\mathrm{del,tot}}(1) \approx 6.6 \times 10^{-4} (m_{\mathrm{pbh}}/M_\odot) \, \mathrm{yr}^{-1}$, and an absolute global lunar impact rate of $\Gamma_{\mathrm{lunar}}^{\mathrm{max}} = 3.18 \times 10^{-5} \, \mathrm{yr}^{-1}$ (derived from the $3$\,Gyr chronological average \cite{Neukum2001}), the inequality becomes:
\begin{equation}
(7.4 \times 10^{-3}) \, f_{\mathrm{PBH}} \left[ 6.6 \times 10^{-4} \left(\frac{m_{\mathrm{pbh}}}{M_\odot}\right) \right] \lesssim (0.1) \times 3.18 \times 10^{-5}.
\end{equation}
Solving this yields the constraint:
\begin{equation}
f_{\mathrm{PBH}} \lesssim 0.65 \left(\frac{m_{\mathrm{pbh}}}{M_\odot}\right)^{-1}.
\end{equation}

Remarkably, this yields a constraint highly consistent with the boundary derived from the terrestrial record. By utilizing the Moon as an independent, erosion-free control volume with its own absolute cratering chronology, this result demonstrates that the severe constraint on stellar-mass PBH dark matter (e.g., $f_{\mathrm{PBH}} \lesssim 6.5 \times 10^{-4}$ for $m_{\mathrm{pbh}}=10^3 M_\odot$) is not an artifact of missing terrestrial geological data. Rather, it is a robust dynamical ceiling imposed by the pristine cratering history of the Earth-Moon system.

\subsection{Combined Limits and Discussion}
\begin{figure*}[t]
\centering
\includegraphics[width=\textwidth,height=0.3\textheight]{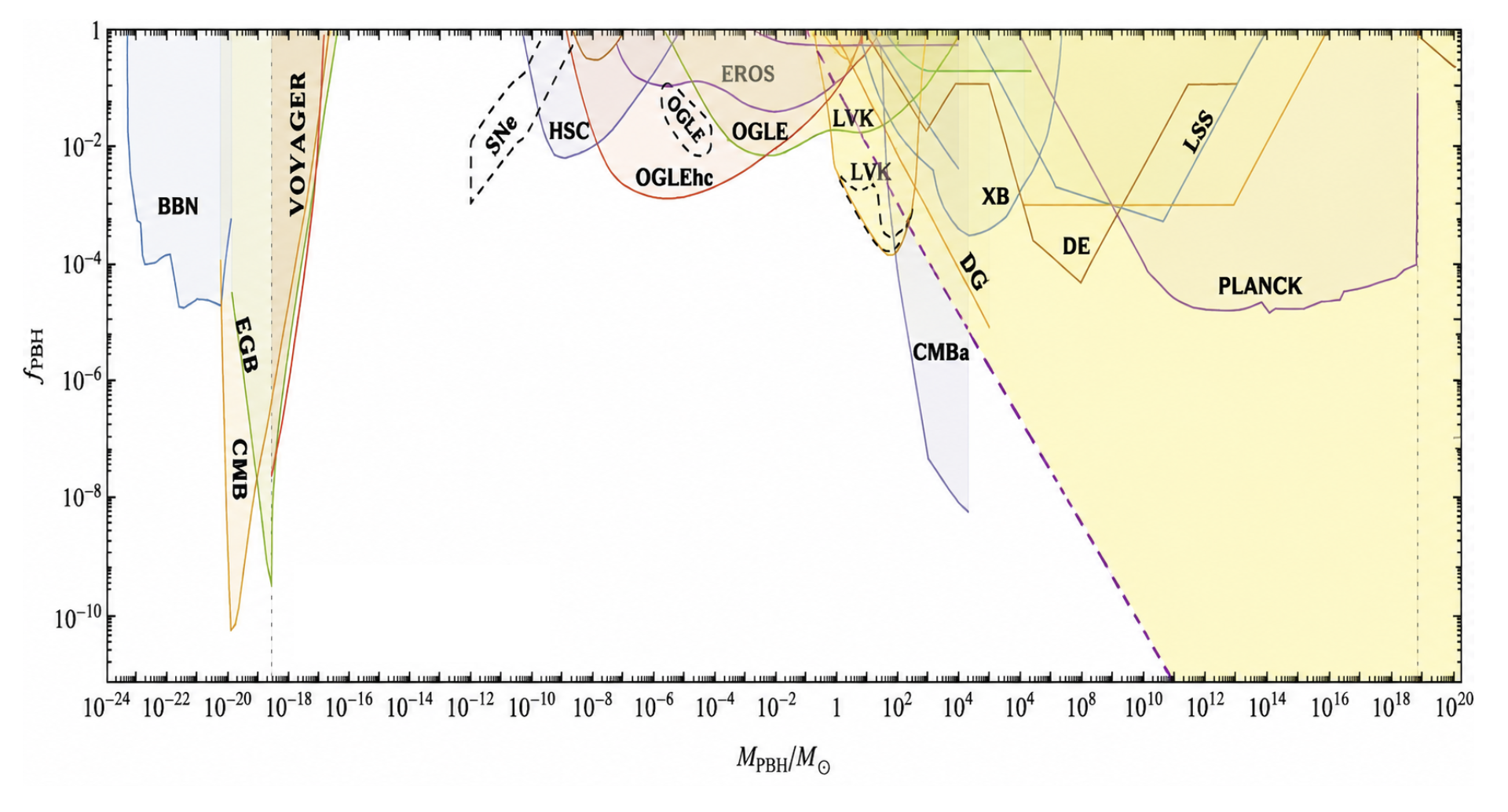}
\caption{Combined constraints on the PBH dark matter fraction, $f_{\mathrm{PBH}}$, from various astrophysical probes, adapted from \cite{2026NCimR.tmp....4C}. The yellow shaded region indicates the exclusion parameter space derived from Oort cloud dynamics and inner Solar System impact records (this work). Our overarching stringent constraint is bounded by $f_{\mathrm{PBH}} \lesssim 0.65 (m_{\mathrm{pbh}}/M_\odot)^{-1}$, effectively excluding intermediate-to-high mass PBHs as a significant dark matter component.}
\label{fig:constraints_combined}
\end{figure*}

Our Oort-cloud-based limits complement and, in the intermediate-to-high mass window ($10^2$–$10^5 M_\odot$), significantly surpass other astrophysical probes. Figure~\ref{fig:constraints_combined} synthesizes our impact constraints with existing limits derived from microlensing surveys (EROS, MACHO, OGLE) \cite{Griest2013, niikura2019, moniez}, Cosmic Microwave Background (CMB) anisotropies \cite{Carr2021}, and wide binary disruptions.

In the mass range above $10^2 M_\odot$, standard constraints face fundamental observational and theoretical limitations. Microlensing bounds weaken considerably because the Einstein crossing times scale as $t_E \propto \sqrt{m_{\mathrm{pbh}}}$, eventually exceeding the baseline durations of modern surveys (typically $\sim 10$ years) for high-mass lenses. Similarly, CMB constraints depend heavily on the complex physics of gas accretion onto PBHs during the cosmic dark ages, making them highly sensitive to assumptions regarding accretion geometry, relative velocity, and radiative efficiency. Wide binary disruption limits, while dynamically robust, remain sensitive to uncertainties in the local dark matter phase-space distribution and the Galactic tidal potential over gigayear timescales.

Our limits rely on the well-calibrated, long-term integration of purely gravitational perturbations on the Oort cloud, corroborated by the pristine cratering history of the Earth-Moon system. Our lunar impact bound provides the most stringent exclusion, establishing an upper limit of $f_{\mathrm{PBH}} \lesssim 0.65 \left(m_{\mathrm{pbh}}/M_\odot\right)^{-1}$. This strictly limits the dark matter fraction to $f_{\mathrm{PBH}} \lesssim 6.5 \times 10^{-4}$ at $10^3 M_\odot$, providing the most robust dynamical constraint in this mass decade. 

\section{Conclusion}
The Oort cloud---a relic of our Solar System's formation---serves as a powerful, gigayear-scale gravitational detector for compact dark matter. By investigating the scattering dynamics between primordial black holes (PBHs) and Oort cloud objects, we have established a novel, Solar System-based method to constrain PBH dark matter across the stellar-to-intermediate mass window.

Our analysis reveals that PBH--Oort object encounter rates follow a linear scaling $\Gamma \propto m_{\mathrm{pbh}}$ for $m_{\mathrm{pbh}} \gtrsim 10^{-10} M_\odot$, leading to observable effects for massive PBHs. Through four independent observational probes---Oort cloud survival, dynamically new long-period comet fluxes, the terrestrial impact rate, and the pristine lunar cratering chronology---we derive stringent upper limits on $f_{\mathrm{PBH}}$. Our most robust combined impact constraints rigorously exclude PBHs as the dominant dark matter component ($f_{\mathrm{PBH}} = 1$) for $10^2 M_\odot \lesssim m_{\mathrm{pbh}} \lesssim 10^5 M_\odot$, establishing a severe upper limit of $f_{\mathrm{PBH}} \lesssim 6.5 \times 10^{-4}$ at $m_{\mathrm{pbh}} = 10^3 M_\odot$. These limits are particularly significant in the $10^2$--$10^4 M_\odot$ range, where they complement and definitively surpass existing microlensing and CMB bounds.

Beyond setting stringent constraints, our work demonstrates that planetary systems can serve as sensitive gravitational laboratories for fundamental physics. The predicted anisotropic signature in comet injection directions---manifesting as a pronounced hemispheric asymmetry in the expected number of comets arriving from the anti-apex versus the apex hemisphere due to the dark matter wind---offers a distinctive, testable prediction that can cleanly differentiate PBH perturbations from standard stellar encounters. Forthcoming deep-sky surveys, particularly the Vera C. Rubin Observatory (LSST), are poised to discover thousands of dynamically new long-period comets, enabling a robust statistical detection of this hemispheric dipole within its first few years of operation.

Looking forward, the synergy between Solar System dynamical detectors, gravitational wave observations from the LIGO-Virgo-KAGRA network, and cosmological probes will continue to illuminate the nature of dark matter. The Oort cloud demonstrates that the long-term dynamical integration of our local galactic environment can provide some of the most profound constraints on the hidden universe.

\begin{acknowledgments}
The author acknowledges the use of an AI‑assisted language for rephrasing the manuscript to comply with the journal’s style. 
\end{acknowledgments}

\bibliography{aipsamp}
\end{document}